\documentclass[10pt,letterpaper]{article}
\usepackage[top=0.85in,left=2.75in,footskip=0.75in]{geometry}

\usepackage{amsmath,amssymb}

\usepackage{changepage}

\usepackage[utf8x]{inputenc}

\usepackage{textcomp,marvosym}

\usepackage{cite}

\usepackage{nameref,hyperref}

\usepackage[right]{lineno}

\usepackage{microtype}
\DisableLigatures[f]{encoding = *, family = * }

\usepackage[table]{xcolor}

\usepackage{array}

\newcolumntype{+}{!{\vrule width 2pt}}

\newlength\savedwidth

\raggedright
\usepackage[aboveskip=1pt,labelfont=bf,labelsep=period,justification=raggedright,singlelinecheck=off]{caption}

\usepackage{todonotes}
\reversemarginpar
\makeatletter
\renewcommand{\@biblabel}[1]{\quad#1.}
\makeatother

\usepackage{lastpage,fancyhdr,graphicx}
\usepackage{epstopdf}
\fancyheadoffset[L]{2.25in}
\fancyfootoffset[L]{2.25in}
\usepackage{xspace} 
\newcommand{\DLC}{fDLC\xspace}

\usepackage{soul}
\usepackage{siunitx}

\begin{document}
\vspace*{0.2in}

\begin{flushleft}
{\Large
\textbf\newline{Fast deep learning correspondence for  neuron tracking and identification in  \textit{C. elegans} using synthetic training}

}





Xinwei Yu\textsuperscript{1},
Matthew S. Creamer\textsuperscript{2},
Francesco Randi\textsuperscript{1},
Anuj K. Sharma\textsuperscript{1},
Scott W. Linderman\textsuperscript{3,4},
Andrew M. Leifer\textsuperscript{1,2,*}
\\
\bigskip
\textbf{1} Department of Physics, Princeton University, Princeton, NJ, United States of America
\\
\textbf{2} Princeton Neuroscience Institute, Princeton University, Princeton, NJ, United States of America
\\\textbf{3} Department of Statistics, Stanford University, Stanford, CA, United States of America
\\\textbf{4} Wu Tsai Neurosciences Institute, Stanford University, Stanford, CA, United States of America

\bigskip

* leifer@princeton.edu

\end{flushleft}
\section*{Abstract}
We present an automated method to track and identify neurons in \textit{C. elegans}, called “fast Deep Learning Correspondence” or fDLC, based on the transformer network architecture. 
The model is trained once on empirically derived synthetic data and then predicts neural correspondence across held-out real animals via transfer learning. The same pre-trained model both tracks neurons across time and identifies corresponding neurons across individuals. Performance is evaluated against hand-annotated datasets, including NeuroPAL \cite{yeminiNeuroPALMulticolorAtlas2020}. Using only position information, the method achieves 80.0\% accuracy at tracking neurons within an individual and 65.8\% accuracy at identifying neurons across individuals. Accuracy is even higher on a published dataset \cite{chaudharyAutomatedAnnotationCell2020}. Accuracy reaches 76.5\% when using color information from NeuroPAL. Unlike previous methods, fDLC  does not require straightening or transforming the animal into a canonical coordinate system. The method is fast and predicts correspondence in 10 ms making it suitable for future real-time applications.



\section*{Introduction}
The nervous system of the nematode \textit{C.elegans} is well characterized, such that each of the 302 neurons are named and have stereotyped locations across animals~\cite{noauthor_structure_1986, sulston_post-embryonic_1976, Witvliet2020.04.30.066209}. The capability to find corresponding neurons across animals is essential to investigate neural coding and  neural dynamics across animals. Despite the worm’s overall stereotypy, the variability in neurons' spatial arrangement is sufficient to make  predicting neural correspondence a challenge. For whole-brain calcium imaging~\cite{venkatachalam_pan-neuronal_2016, nguyen_whole-brain_2016}, identifying neurons across animals is additionally challenging because the nuclear localized markers  that are used tend to obscure morphological features that would otherwise  assist in neural identification. 

An ideal  method for finding neural correspondence in \textit{C. elegans} should accommodate  two major use cases. The first is tracking neurons within an individual across time as the animal's head moves and deforms. Here the goal is to be able to say with confidence that a neuron imaged in a volume taken at time $t_1$ is the same as another neuron taken from a volume imaged at time $t_2$.  Tracking across time is needed to extract  calcium dynamics from neurons  during freely moving population calcium imaging~\cite{venkatachalam_pan-neuronal_2016, nguyen_whole-brain_2016, lagacheRobustSingleNeuron2020}. Additionally, very fast real-time tracking will be needed to guide closed-loop  optogenetic stimulation of neurons in moving animals as those methods move to larger neural populations~\cite{leiferOptogeneticManipulationNeural2011, stirmanRealtimeMultimodalOptical2011, kocabasControllingInterneuronActivity2012, shipleySimultaneousOptogeneticManipulation2014}.    

The second and more general use case is finding neural correspondence across individuals. Often this is to identify the name of a neuron  with respect to the connectome~\cite{noauthor_structure_1986} or a gene expression atlas \cite{hammarlundCeNGENProjectComplete2018}. Even when a neuron's name cannot be ascertained, being able to identify which neurons are the same across recordings allows researchers to study neural population codes common across individuals.

For both use cases, a method to find neural correspondence is desired that is accurate, fast, requires minimal experimental training data and that  generalizes across animal pose, orientation, imaging hardware, and conditions. 
Furthermore, an ideal method should  not only  perform well when restricted to neural positioning information but, should also be flexible enough to leverage   genetically encoded color labeling information or  other features for improved accuracy when available. Multicolor strains are powerful new tools that use multiple  genetically encoded fluorescent labels to aid  neural identification \cite{yeminiNeuroPALMulticolorAtlas2020, toyoshimaAnnotationDatasetFacilitates2019} (we use one of those strains,  NeuroPAL~\cite{yeminiNeuroPALMulticolorAtlas2020}, for validating our model). However, some applications, like whole-brain imaging in moving worms,  are not yet easily compatible with the multicolor imaging required by these new strains, so  there remains a need for improved methods that use position information alone.

A variety of  automated methods for \textit{C. elegans}  have been developed that address some, but not all of these needs. Most methods developed so far focus on finding the extrinsic similarity \cite{bronsteinRockPaperScissors2007} between one neuron configuration, called a test, and another neuron configuration called a template.   
Methods like these  deform space  to minimize  distances between  neurons in the template  and  neurons in the test  and then attempt to solve an assignment problem \cite{lagacheTrackingActivityDeformable2018}.  For example, a simple implementation would be to use a non-rigid  registration model, like Coherent Point Drift (CPD) \cite{myronenko_point_2010} to optimize a  warping function between neuron positions in the test  and template. More recent non-rigid registration algorithms like PR-GLS \cite{PR-GLS2016} also  incorporate  relative spatial arrangement of the neurons  \cite{Wen385567}.

Models can also do better by incorporating the statistics of neural variability. 
NeRVE registration and clustering \cite{nguyenAutomaticallyTrackingNeurons2017}, for example, also uses a non-rigid point set registration algorithm \cite{jianRobustPointSet2011} to find a warping function that minimizes the difference between a configuration of neurons at one time point and another. But  NeRVE  further registers the test neurons onto multiple templates to define a feature vector and then  finds neural correspondence by clustering those feature vectors. By using  multiple templates, the method implicitly incorporates more information about the range and statistics of that individual animal’s poses to improve accuracy. 

A related line of work uses generative models to capture the statistics of variability across many individual worms. These generative models  specify a joint probability distribution over neural labels and the locations, shapes, sizes, or appearance of neurons identified in the imaging data of multiple individuals~\cite{bubnisProbabilisticAtlasCell2019, Varol2020-xo, Nejatbakhsh2020-pr, nejatbakhshNeuronMatchingElegans2021}.  
These approaches are based on assumptions about the likelihood of observing a test neural configuration, given an underlying configuration of labeled neurons. For example, these generative models often begin with a Gaussian distribution over neuron positions in a canonical coordinate system and then assume a distribution over potentially non-rigid transformations of the worm's pose for each test configuration. Then, under these assumptions, the most likely neural correspondence is estimated via approximate Bayesian inference.

The success of generative modeling hinges upon the accuracy of its underlying assumptions, and these are challenging to make for high-dimensional data. 
An alternative is to take a discriminative modeling approach~\cite{Bishop2006-tx}. For example, recent work~\cite{chaudharyAutomatedAnnotationCell2020} has used conditional random fields (CRF) to directly parameterize a conditional distribution over neuron labels, rather than assuming a model for the high-dimensional and complex image data.
CRF allows for a wide range of informative features to be incorporated in the model, such as the angles between neurons, or their  relative anterior-posterior positions, which are known to be useful for identifying neurons~\cite{long3DDigitalAtlas2009}. Ultimately, however, it is up to the modeler to select and hand curate a set of features to input into the CRF.

The next logical step is to allow for much richer features to be learned from the data. Artificial neural networks are ideal for tackling this problem, but they require immensely large training sets. Until now, their use for neuron identification has been limited. For example, in one tracking algorithm,  artificial neural networks  provide only the initialization, or first guess, for non-rigid registration \cite{Wen385567}. 

Our approach is based on a simple insight: it is straightforward to generate very large synthetic datasets of test and template worms that nonetheless are derived from measurements. We use neural positions extracted from existing imaging datasets, and then apply known, nonlinear transformations to warp those positions into new shapes for other body postures.  Furthermore, we simulate the types of noise that appear in real datasets, such as missing or spurious neurons. Using these large-scale synthetic datasets, we can train an artificial neural network to map the simulated neural positions back to the ground truth.  
Given sufficient training data (which we can generate at will), the network learns the most informative features of the neural configurations, rather than requiring the user to specify them by hand.
Realistic synthetic or augmented datasets like these have been key to cracking other challenging problems in neural and behavioral data analysis~\cite{lee2020yass,mathis2020deep}, and have  already shown promising potential for tracking neurons \cite{Wen385567}.

In this work, we use synthetic data to train a Transformer network, an artificial neural network architecture that has shown great success in natural language processing tasks~\cite{bibTransformer}. 
Transformers incorporate an attention mechanism that can leverage similarities between pairs of inputs to build a rich representation of the input sequence for downstream tasks like machine translation and sentiment prediction. We reasoned this same architecture would be well-suited to   extract spatial relationships between neurons in order to build a representation that facilitates finding  correspondence to neurons in a template worm. 

Not only is the Transformer well-suited to learning features for the neural correspondence problem, it also obviates the need to straighten \cite{pengStraighteningCaenorhabditisElegans2008} the worm in advance.  Until now, existing methods have either required the worm to be straightened in preprocessing~\cite{bubnisProbabilisticAtlasCell2019, chaudharyAutomatedAnnotationCell2020} or explicitly transformed them during inference~\cite{Varol2020-xo, Nejatbakhsh2020-pr}. Straightening the worm is a non-trivial task, and it is especially error-prone for complicated poses such as when the worm rolls along its centerline.

Finally, one of the main advantages of the Transformer architecture is that it permits parallel processing of the neural positions using modern GPU hardware. In contrast to existing methods, which have not been optimized for speed, the Transformer can make real-time predictions once it has been trained. This speed is a necessary step toward future experiments with real-time, closed-loop, targeted delivery of optogenetic stimulation of large populations of neurons in freely moving animals.

\section*{Materials and methods}
\subsection*{Datasets}
Our model was trained on a synthetic dataset derived from recordings of  12 freely moving animals. The model's performance was evaluated on a different set of 12 held-out recordings:  one moving recording and 11 immobile recordings. The recording of moving animals used strain  AML32  wtfIs5[P\textit{rab-3}::NLS::GCaMP6s; P\textit{rab-3}::NLS::tagRFP]  and was first reported in ~\cite{nguyenAutomaticallyTrackingNeurons2017}.  Immobile recordings to evaluate performance used strain  AML320 (otIs669[NeuroPAL] V 14x; wtfIs145 [pBX + rab-3::his-24::GCaMP6::unc-54]
) derived from NeuroPAL strain  OH15262 ~\cite{yeminiNeuroPALMulticolorAtlas2020}.  All strains and datasets in this study used nuclear localized fluorescent reporters.

Neural configurations acquired as part of this study have been posted in an Open Science Foundation repository with DOI:10.17605/OSF.IO/T7DZU available at \url{https://dx.doi.org/10.17605/OSF.IO/T7DZU}.

Model performance was also evaluated  on a publicly accessible dataset from  \cite{chaudharyAutomatedAnnotationCell2020} available at \url{https://github.com/shiveshc/CRF_Cell_ID}, commit \texttt{74fb2feeb50afb4b840e8ec1b8ee7b7aaa77a426}.

\subsection*{Imaging}
To image neurons in the head of freely moving worms, we used a dual-objective spinning-disk based tracking system  ~\cite{nguyen_whole-brain_2016} (Yokogawa CSU-X1 mounted on a Nikon Eclipse TE2000-S). Fluorescent images of the head of a worm were recorded through a 40x objective with both 488- and 561-nm excitation laser light as the animal crawled. The 40x objective  translated up and down along the imaging axis to acquire 3D image stacks at a rate of 6 head  volumes/s.

To image neurons in the immobile multi-color NeuroPAL worms ~\cite{yeminiNeuroPALMulticolorAtlas2020} we modified our setup by adding 
emission filters in a motorized filter wheel (Prior ProScan-II), and adding 
a Stanford Research Systems SR474 shutter controller (with SR475 shutters) to programmatically illuminate the worm with  different wavelength laser light. We use three lasers with light of different wavelengths: 405 nm (Coherent OBIS-LX 405 nm 100 mW), 488 nm (Coherent SAPPHIRE 488 nm 200 mW), and 561 nm (Coherent SAPPHIRE 561 nm 200 mW). Only one laser at a time reached the sample, through a 40x oil-immersion objective (1.3 NA, Nikon S Fluor). The powers measured at the sample, after spinning disk and objective, were 0.14 mW (405 nm), 0.35 mW (488 nm), and 0.36 mW (561 nm). In the spinning disk unit, a dichroic mirror (Chroma ZT405/488/561tpc) separated the excitation from the emission light. The latter was relayed to a cooled sCMOS camera (Hamamatsu ORCA-Flash 4.0 C11440-22CU), passing through the filters mounted on the filter wheel (Table \ref{table:methods:filters}). Fluorescent images were acquired in different “channels”, i.e. different combinations of excitation wavelength, emission filter, and camera exposure time (Table \ref{table:methods:channels}). The acquisition was performed using a custom software written in LabVIEW that specifies the sequence of channels to be imaged, and controls shutters, filter wheel, piezo translator, and camera. After setting the z position, the software acquires a sequence of images in the specified channels. 

\begin{table}
\centering
\caption{
{\bf List of emission filters for multicolor imaging.}}
\begin{tabular}{c|c}
Filter label & Filters (Semrock part n.)    \\
\hline
F1 & FF01-440/40                            \\
F2
& FF01-607/36                            \\
F3
& FF02-675/67 + FF01-692/LP              \\
\end{tabular}
\label{table:methods:filters}
\end{table}

\begin{table}
\centering
\caption{
{\bf Imaging channels used.}}
\begin{tabular}{c|ccc}
Channel & Excitation $\lambda$ (nm) & Emission window (nm) [filter]  & Primary Fluorophore \\
\hline
ch0 & 405 & 420-460 [F1] & mtagBFP \\
ch1 & 488 & 589-625 [F2] & CyOFP      \\
ch2 & 561 & 589-625 [F2] & tagRFP-t  \\
ch3 & 561 & 692-708 [F3] & mNeptune \\
\end{tabular}
\label{table:methods:channels}
\end{table}

\subsection*{Preprocessing and Segmentation}
We extracted the position of individual neurons from  3D fluorescent  images to generate a 3D point cloud,  (Fig.~\ref{fig1}A).  This process is called segmentation and the \DLC model is agnostic to the specific choice of the segmentation algorithm. 

For recordings of strains AML32, we used a segmentation algorithm adopted from  ~\cite{nguyenAutomaticallyTrackingNeurons2017}. We first applied a threshold to find pixels where the intensities are significantly larger than the background. Then, we computed the 3D Hessian matrix and its eigenvalues of the intensity image. Candidate neurons were regions where the maximal eigenvalue was negative. Next, we searched for the local intensity peaks in the region and spatially disambiguated peaks in the same region with a watershed separation based on pixel intensity. 

For recordings of NueroPAL strains, we used the same segmentation algorithm as in \cite{yeminiNeuroPALMulticolorAtlas2020}.   The publicly accessible dataset from \cite{chaudharyAutomatedAnnotationCell2020} used in Fig.~\ref{fig4} had already been segmented prior to our use.

\subsection*{Generating synthetic point clouds for training}
We developed a simulator to generate a large training set of synthetic animals  with known neural correspondence. 
The simulator takes as its input the point clouds collected from approximately 4,000 volumes spread across recordings of 12 freely moving animals. For each volume, the simulator performs a series of stochastic deformations and transformations to generate 64 new synthetic individuals where the ground truth correspondence  between neurons in the individuals and the original point cloud is known.  A total of \num{2.304e5} synthetic point clouds were used for training.  

The simulator introduces a variety of different sources of variability and   real-world deformations to create each synthetic point cloud  (Fig.~\ref{fig1}B,E). The simulator starts by straightening the worm in the XY plane using its centerline so that it now lies in a canonical worm coordinate system. Before straightening, Z is along the optical axis and XY are defined to be perpendicular to the optical axis and are arbitrarily set by the orientation of the camera.  After straightening, the animal's posterior-anterior axis lies along the X axis.
To introduce animal-to-animal variability in relative neural position, a non-rigid transformation is applied to the neuron point cloud against a template randomly selected from recordings of the real observed worms using coherent point drift (CPD)  ~\cite{myronenko_point_2010}.  To add variability associated with  rotation and distortion of the worm's head in the transverse plane, we apply a random affine transformation to the transverse plane.  To simulate missing neurons and segmentation errors, spurious neurons are randomly added, and some true neurons are randomly removed, for up to 20\% of the observed neurons.  To introduce variability associated with animal pose, we randomly deform the centerline of the head.  Lastly,  to account for variability in animals' size and orientation, a random affine transformation in XY plane is applied that rescaled the animal's size by up to 5\%. With those steps, the simulator deforms a sampled worm and generates a new synthetic worm with different orientation and posture while maintaining known correspondence.   

Whenever practical, the magnitude of the simulator's perturbations was informed by observation of real worms. For example the centerlines generated by the simulator were directly sampled from recordings of real individuals. The magnitude of added Gaussian noise was set to have a  standard deviation of $0.42\mu m$, as this roughly matched our estimate of variability observed by eye.

\subsection*{Deep Learning Correspondence Model}
\subsubsection*{Overview and input}
The deep learning correspondence model (\DLC) is an artificial neural network based on the   Transformer~\cite{bibTransformer} architecture (Fig.~\ref{fig1}C) and is implemented in the automatic differentiation framework PyTorch~\cite{Paszke2017AutomaticDI}. The \DLC model takes as input the positional coordinates of a pair of worms, a template worm $a$, and test worm, $b$ (Fig.~\ref{fig1}F). For each worm, approximately 120 neurons are segmented and passed to the \DLC model. 

\subsubsection*{Architecture}
The model works as an encoder, which maps the input neuron coordinates $(a_1, a_2, ..., a_n, b_1, b_2, ..., b_m)$ to continuous embeddings $(u_1, u_2, ...u_n, v_1, v_2, ..., v_m)$.
The model is composed of a stack of $N=6$ identical layers. Each layer consists of two sub-layers: a multi-head self-attention mechanism~\cite{bibTransformer}, and a fully connected feed-forward network.   The multi-head attention mechanism is  the defining feature of the transformer architecture and makes the architecture well-suited for finding relations in sequences of data, such as words in a sentence or, in our case, spatial locations of neurons in a worm.   Each head contains a one-to-one mapping between the nodes in the artificial network and the \textit{C. elegans} neurons.  In the transformer architecture, features of a  previous layer are mapped via a linear layer into three attributes of each node, called the  query, the  key and the value pairs. These attributes of each node contain high dimensional feature vectors which, in our context,  represent information about the neuron's relative position. The multi-head attention mechanism computes a weight for each pair of nodes (corresponding to each pair of \textit{C. elegans} neurons). The weights are calculated by performing a set computation on the query and  key. The output is calculated by multiplying this resultant weight by the value. In our implementation, we set the number of heads in the multi-head attention module to be 8 and we set the dimension of our feature vectors to be 128. We choose the best set of the hyperparameters by evaluating on a validation set, which is distinct from the training set and also from any data used for evaluation. A residual connection~\cite{he_deep_2016} and layer normalization~\cite{ba2016layer} are employed for each sub-layer, as is widely used in artificial neural networks.

\subsubsection*{Calculating probabilities for potential matches}
The \DLC model generates a high dimensional ($d=128$) embedding $u_i$ for neuron $i$ from the template worm and $v_j$ for the neuron $j$ from the test worm. The similarity of a pair of embeddings, as measured by the inner product~$\langle u_i, v_j \rangle$, determines the probability that the pair is a match. Specifically, we define the probability that neuron~$i$ in the template worm matches  neuron $j$ in the test worm as $p_{ij}$, where
\begin{eqnarray}
\label{eq:Pij}
    p_{ij} &= \frac{e^{\langle u_i, v_j\rangle}}{\sum_{k=1}^m e^{\langle u_i, v_k \rangle}}.
\end{eqnarray}
Equivalently, the vector~$p_i = (p_{i1}, \ldots, p_{im})$ is modeled as the ``softmax'' function of the inner products between the embedding of neuron~$i$ and the embeddings of all candidate neurons~$1,\ldots,m$. 
The softmax output is non-negative and sums to one so that $p_{i}$ can be interpreted as a discrete probability distribution over assignments of neuron~$i$. 

We also find the most probable correspondence between the two sets of neurons by solving a maximum weight bipartite matching problem where the weights are given by the inner products between test and template worm embeddings. This is a classic combinatorial optimization problem, and it can be solved in polynomial time using the Hungarian algorithm~\cite{kuhn_hungarian_1955}. 

\subsubsection*{End-user output}
The \DLC model returns two sets of outputs to the end user. One is the algorithm's estimate of the most probable labels for each neuron in the test worm; i.e. the solution to the maximum weight bipartite matching problem described above. The other is an ordered list of alternative candidate labels for each individual neuron in the test worm and their probabilities  ranked from most to  least probable.

\begin{figure}[htbp]
 	\includegraphics[width=0.95\linewidth]{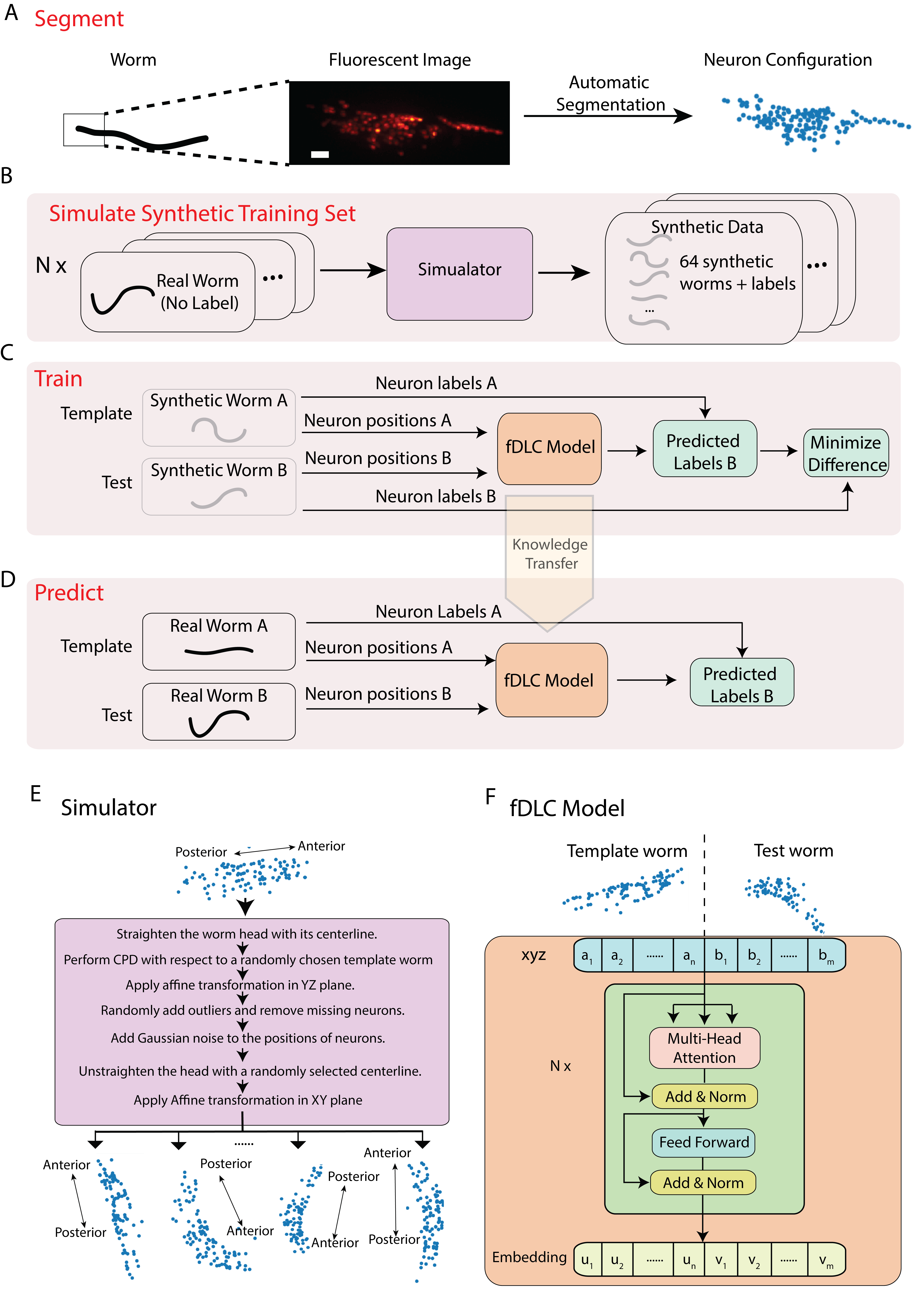}
\caption{{\bf Fast Deep Learning Correspondence Model.} (Caption continued on the next page.)}
\label{fig1}
\end{figure}

\addtocounter{figure}{-1}
\begin{figure}[h]
\caption{
{\bf Fast Deep Learning Correspondence Model}
(A-D) schematic of training and analysis pipeline for using the fast Deep Learning Correspondence (\DLC) model to predict  correspondence between neurons across  individuals. (A) Volumetric images of fluorescent labeled neuronal nuclei are segmented to extract neuron positions.  (Scale bar, $10\mu m$). (B) Synthetic training data is generated with a simulator. The simulator  transforms the neural positions of a real worm and introduces noise to generate new synthetic individuals.
Approximately $N=10^4$ neuron configurations without labels from 12 moving worms were used to generate  \num{2.304e5} labeled synthetic worms  for training. (C)  During training,  the \DLC model finds optimal internal parameters to minimize the difference between predicted neural labels and true labels in pairs of synthetic worms. 
(D) Given  position and labels for neurons in real worm A and  position for neurons in real worm B, the trained model predicts corresponding labels for worm B. (E) Detailed schematic of the simulator from panel B.   (F) Transformer architecture of the \DLC model. The position features of a template worm with $n$ neurons and a test worm with $m$ neurons are taken as input. The features are computed via a multi-head attention mechanism. `Add \& Norm' refers to an addition and layer normalization step. $a$ and $b$ are neuron positions and $u$ and $v$ are embeddings  for the template and test, respectively. We choose the number of layers $N=6$ and the  embedding dimension $d_{emb}=128$ by evaluating the performance on a held-out validation set.}
\end{figure}

\subsubsection*{Training}
The model was trained on \num{2.304e5} synthetic animals derived from  recordings of 12 individuals. The model was trained only once and the same trained model was used throughout this work. 

Training is as follows. We performed supervised learning with ground truth labels of neuron names provided by the synthetically generated data. A cross-entropy loss function was used. If neuron $i$ and neuron $j$ has the same name, the cross-entropy loss function favors the model to output $p_{ij} = 1$. If neuron $i$ and neuron $j$ have different names, the loss function favors the model to output $p_{ij} = 0$. 
The model was trained for 12 hours on a 2.40 GHz Intel machine with NVIDIA Tesla P100 GPU.

\subsection*{Evaluating model performance and comparing against other models}

\subsubsection*{Minimum confidence threshold}
Accuracy of the \DLC model is  calculated by including all neural matches between the test and template regardless of confidence, except for in the case of Fig.~\ref{fig3}. In that analysis only, we sought to reduce the number of matches to better compare to the NeRVE method which only matches 80\% of the neurons. We therefore excluded matches below  a minimum confidence threshold of 0.05 from our calculation of accuracy. The tradeoff between accuracy and coverage is explored further in Fig.~\ref{fig4}G.

\subsubsection*{Coherent Point Drift}
We use Coherent Point Drift (CPD) Registration~\cite{myronenko_point_2010}  as a baseline with which to compare our model's performance.  In our implementation, CPD is used to find the optimal non-rigid transformation to align the test worm with respect to the template worm. We then calculated the distance for each pair of the neurons from the transformed test worm and the template worm. We used the Hungarian algorithm~\cite{kuhn_hungarian_1955} to find the optimal correspondence that minimizes the total squared distance for all matches.

\subsection*{Color Model}
The recently developed NeuroPAL strain ~\cite{yeminiNeuroPALMulticolorAtlas2020} expresses four different genetically encoded fluorescent proteins in specific expression patterns so that a  human can better identify neurons across animals.  Manual human annotation based on these expression patterns serves as ground truth in evaluating our model's performance throughout this work. In Fig.~\ref{fig5}B we also explored combining  color information with our \DLC model. To do so we developed a simple color matching model that operated in parallel to our position based \DLC model. Outputs of both models were then combined to predict the final correspondence between neurons.  

Our color matching model consists of two steps: First, the intensity of each of the color channels is normalized by the total intensity. Then the similarity of color for each pair of neurons is measured as the inverse of the Kullback–Leibler divergence between their normalized color features. 

To  calculate the final combined matching matrix, we add the color similarity matrix  to the position matching log probability matrix from our \DLC model. The similarity matrix of color is multiplied by a factor $\lambda$. We chose $\lambda=60$ so that the amplitude of values in the similarity matrix of color is comparable to our \DLC output. 
We note the matching results are not particularly sensitive to the choice of $\lambda$. The most probable matches are obtained by applying Hungarian algorithm on the combined matching matrix. 

\subsection*{Code}
Source code in Python is provided for the model, for the simulator, and for training and evaluation. A jupyter notebook with a simple example is also provided.  Code is available at \url{https://github.com/XinweiYu/fDLC_Neuron_ID}

\section*{Results}

\subsection*{Fast deep learning correspondence  accurately labels neurons across synthetic individuals }

We developed a fast deep learning correspondence (\DLC) model that seeks to find the correspondence between  configurations of \textit{C. elegans} neurons in different individuals or in the same individual across time (Fig.~\ref{fig1}).  
We used a deep learning artificial neural network architecture, called the transformer architecture \cite{bibTransformer}, that specializes at finding pairs of relations in datasets. The transformer architecture identified similarities across spatial relations of neurons in a test and a template to identify  correspondences between the neurons. 

Within a single individual, neural  positions vary  as the worm moves, deforms, and changes its orientation and pose. Across  isogenic individuals, there is an additional source of variability that arises from  the animal's  development.  In practice, further variability also arises from experimental measurements: neuron positions must first be extracted from  fluorescent images, and  slight differences in label expression, imaging artifacts, and optical scattering all contribute to errors in segmenting individual neurons. 

We created a simulator to model these different sources of variability and used it to generate realistic pairs of empirically derived synthetic animals with known  correspondence between their neurons for training our model (Fig.~\ref{fig1}B, E). The simulator took configurations of neuron positions from  real worms as inputs and then scaled and deformed them, forced them to adopt  different poses sampled from real worms, and then introduced additional sources of noise to generate many new  synthetic individuals. We then trained our \DLC model on these experimentally derived synthetic individuals of different sizes and poses. 

Training our model on the empirically derived synthetic data offered  advantages compared to experimentally acquired data. First, it allowed us to train on larger datasets than would otherwise be practical. We trained  on \num{2.304e5} synthetic individuals, but only  seeded our simulator with unlabeled neural configurations from experimentally acquired recordings of  12 individuals. 
Second, we did not need to provide ground truth correspondence because the simulator instead generates its own ground truth correspondence between synthetic individuals, thereby avoiding a tedious and error prone manual step.  We note we still  use  human annotation for validation, but this requires much smaller data sets.
Third, by using large and varied synthetic data, we force the model to generalize its learning to a wide range of variabilities in neural positions and we avoid the risks of overtraining  on idiosyncrasies  specific to our imaging conditions or segmentation. Overall, we reasoned that training with synthetic data should make the model more robust and more accurate across a wider range of conditions, orientations and animal poses than would be practical with experimentally acquired datasets.

\begin{figure}[htbp]
 	\includegraphics[width=0.9\linewidth]{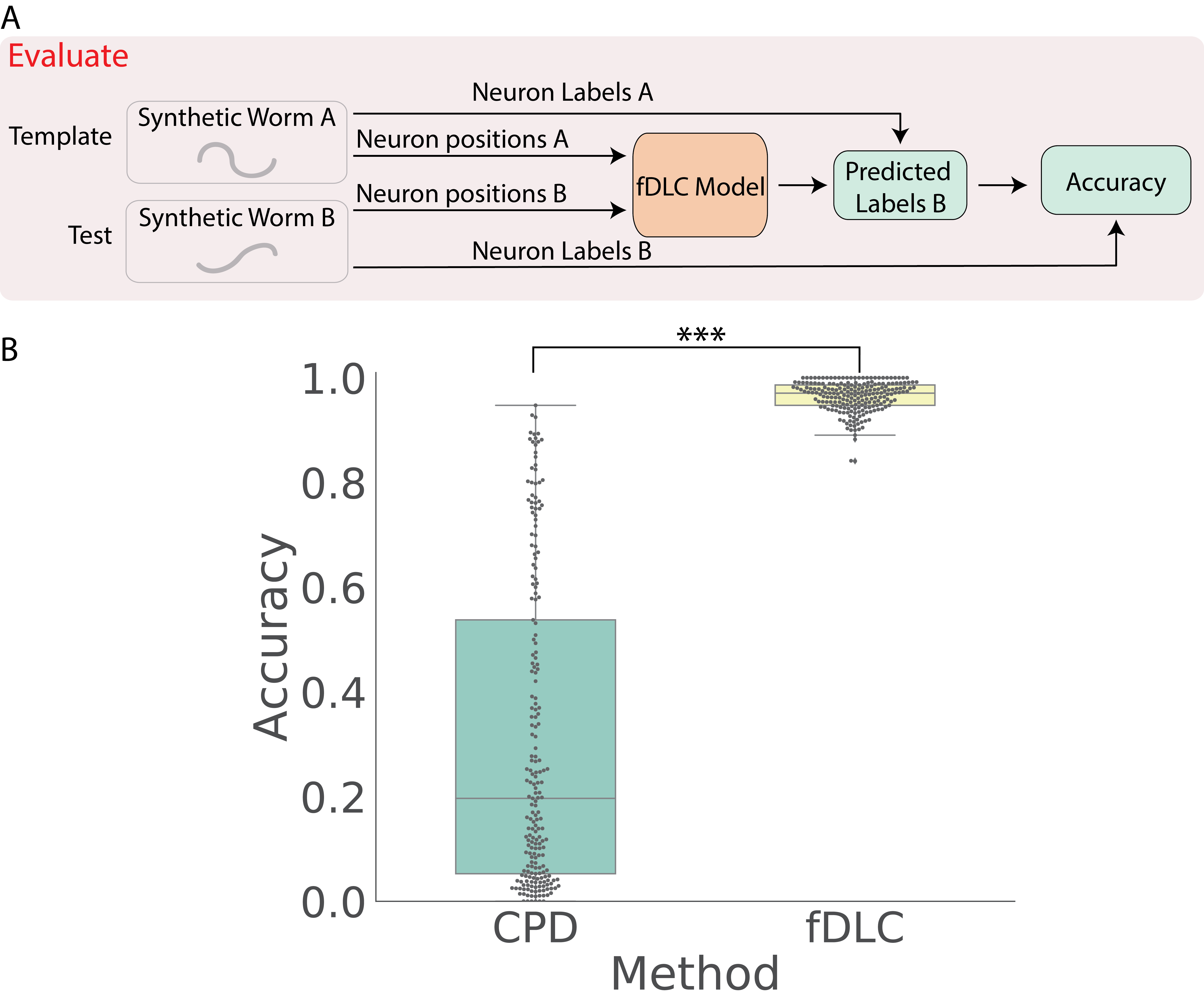}
\caption{{\bf \DLC accurately predicts neuron labels of synthetic worms}
(A) Schematic of evaluation pipeline. \DLC model performance is evaluated on  pairs of synthetic worms with known labels that had been held out from training.  Given neural positions in  worms A and B, and neuron labels from A, the model predicts neuron labels for B. Accuracy is the percent of  labeled neurons, present in both A and B, that the model correctly predicts. (B) Model performance of a Coherent Point Drift Registration (CPD) is compared to the \DLC model on 200 randomly selected pairs of held-out synthetic individuals, without replacement. ($p=\num{2.35e-37}$, Wilcoxon signed rank test).}
\label{fig2}
\end{figure}

We trained our \DLC model on \num{2.304e5} synthetic individuals (Fig.~\ref{fig1}C and methods) and then evaluated its performance on 200 additional held-out synthetic pairs of individuals which had not been accessible to the model during training  (Fig.~\ref{fig2}). 
Model performance was evaluated by calculating the accuracy of the models' predicted correspondence with respect to the ground truth  in pairs of synthetic individuals. One individual is called  the ``test'' and the other is the ``template''.  Accuracy is reported as the fraction of neurons, present in both the test and the template, that were correctly matched.  Our \DLC model achieved 96.6\% average accuracy on the 200 pairs of held-out synthetic individuals.  We compared this performance to that of Coherent Point Drift (CPD)~\cite{myronenko_point_2010}, a classic registration method used for automatic cell annotation.  CPD achieved 30.8\% mean accuracy on the same held-out synthetic individuals. Our measurements show that the \DLC model significantly outperforms CPD at finding correspondence in synthetic data.  For the rest of the work, we use experimentally acquired human annotated data to evaluate performance.


\subsection*{\DLC  accurately tracks neurons within an individual across time}
\begin{figure}[htbp]
 	\includegraphics[width=0.9\linewidth]{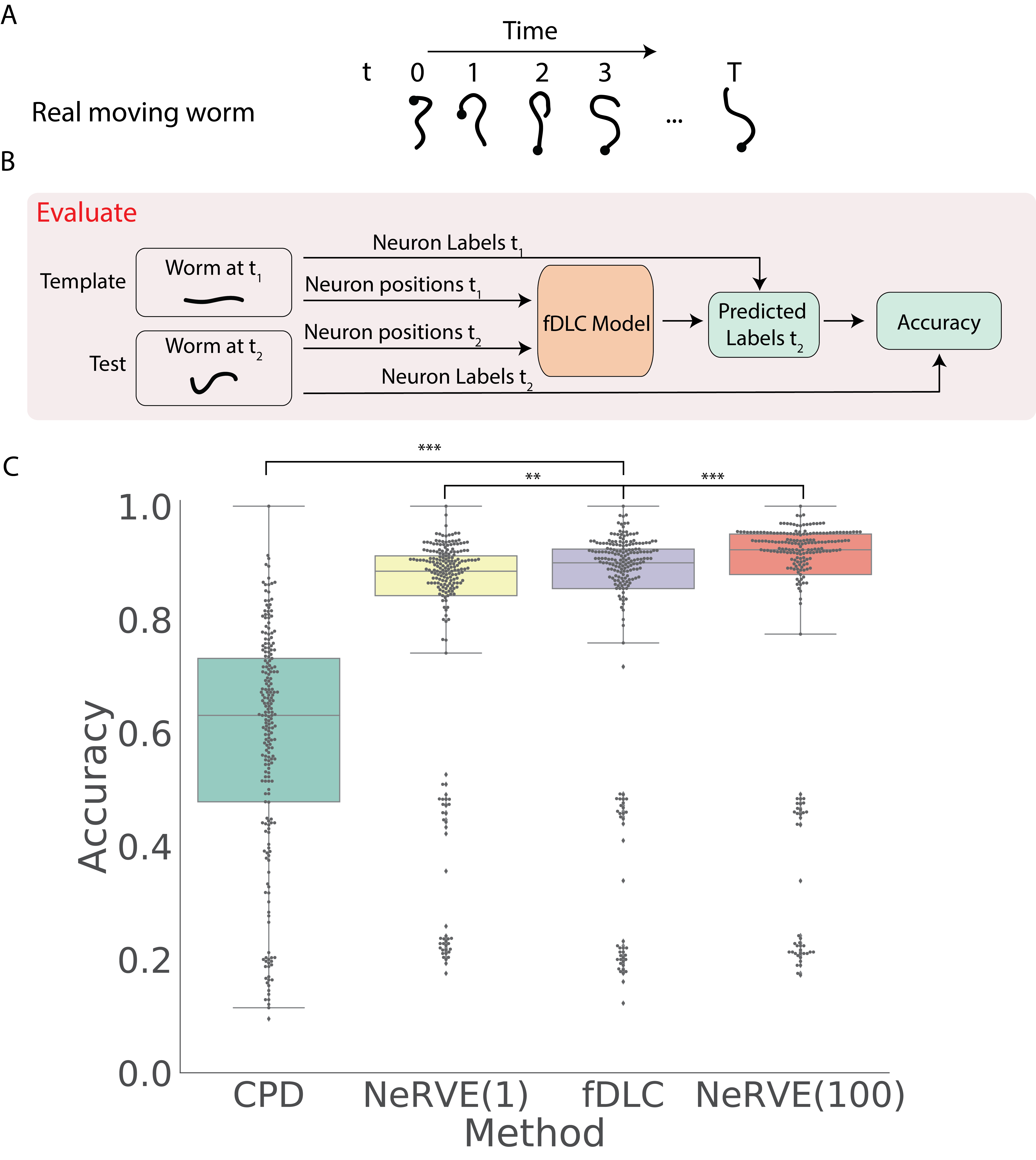}
\caption{{\bf Tracking neurons within an individual across time.}
(A) Schematic shows how the pose and orientation of a freely moving animal change with time. Black dot indicates head.  
(B) Pipeline to evaluate the \DLC model at tracking  neurons within an individual across time. The \DLC model takes in  positional features of a template neuron configuration from one time $t_1$ of a freely moving worm, and predicts the correspondence at another time  $t_2$, called the test. Recording is from \cite{nguyenAutomaticallyTrackingNeurons2017}. Ground truth neuron labels are provided by manual human annotation.  The same time point is used as the  template for all 200 template-test pairs.
(C)  Performance of \DLC and alternative models at tracking neurons within an individual are displayed in order of mean performance. CPD refers to Coherent Point Drift.  NeRVE(1) refers to the restricted NeRVE model that has access to only the same template as CPD and \DLC. NeRVE(100) refers to the full NeRVE model which uses 100 templates from the same individual to make a single prediction. A Wilcoxon signed rank significance test of  \DLC's performance compared to CPD, NeRVE(1) and NeRVE(100) yields $p=\num{4.9e-37}, \num{0.006}$ and $\num{3.1e-18}$ respectively. Boxplots show median and interquartile range.}
\label{fig3}
\end{figure}

We next   evaluated the \DLC model's performance at tracking neurons within an individual over time,  as is needed, for example, to measure calcium activity in moving animals ~\cite{venkatachalam_pan-neuronal_2016, nguyen_whole-brain_2016}. We evaluated model performance on an experimentally acquired calcium imaging recording of a freely moving \textit{C. elegans} (strain AML32) from \cite{nguyenAutomaticallyTrackingNeurons2017}  in which a team of human experts had manually tracked and annotated neuron positions over time. This recording had been excluded from the set of recordings used by the simulator.  We sampled neuron configurations from 201 different time points during this recording to form 200 pairs of  configurations upon which to evaluate the \DLC model. Each pair consisted of a test and template.  The template was always from the same time point $t$,  while the  test was taken to be any of the other 200 time points. 
We applied the pre-trained \DLC model to the 200 pairs of neuron configurations  and compared the model's predicted correspondence to  the ground truth from manual human tracking (Fig.~\ref{fig3}). Across the 200 pairs, the \DLC model showed an average accuracy of 80.0\%. We emphasize that the \DLC model achieved this high accuracy on tracking a real worm using only neuron position information even though it is  trained exclusively on  synthetic data. 

We compared the performance of  our  \DLC model to that of  CPD Registration, and to Neuron Registration Vector Encoding  and clustering (NeRVE), a classical computer vision model that we had previously developed specifically for tracking neurons within a  moving animal over time  \cite{nguyenAutomaticallyTrackingNeurons2017} (Fig.~\ref{fig3}B). NeRVE only matches approximately 80\% of neurons, so for the sake of comparison, in this analysis (but not others), we only consider comparable number of matches from the  \DLC, see ``minimum confidence threshold'' in  ``Materials and Methods.'')
\DLC clearly outperformed CPD achieving  80.0\% accuracy compared to CPD's  58.5\%.  

Both CPD and \DLC  predict neural correspondence of a test configuration by comparing only to a single template. In contrast, the NeRVE method takes   100  templates, where each one is a different  neuron configuration from the same individual,  and uses them all to inform its prediction. The additional templates give the NeRVE method  extra information about the range of possible neural configurations made by the specific individual whose neurons are being tracked.
We therefore compared the \DLC model both to the full NeRVE method and also to a restricted version of the NeRVE method in which  NeRVE had  access only to the same single template as the \DLC model.  (Under this restriction the NeRVE  method no longer clusters and the method collapses to a series of gaussian mixture model registrations \cite{jianRobustPointSet2011}.)  In this way, we could compare the two methods  when given the same information. 
\DLC's mean performance of 80.0\%  was very similar but statistically  significantly more accurate than  the restricted NeRVE model   (mean 79.1\%, $p=6.4\times 10^{-3}$, Wilcoxon signed rank test). The full NeRVE model that had access to  additional templates  outperformed  the \DLC model slightly (82.0\% $p=3.1\times 10^{-18}$, Wilcoxon signed rank test). We conclude that the \DLC model is suitable for tracking individual neurons across time and performs similarly to the NeRVE method. 

In the following sections, we further show that the \DLC method is orders of magnitude faster than NeRVE. Moreover, unlike NeRVE which can only be used within an individual,  \DLC is also able to predict the much more challenging neural correspondence across individuals.

\subsubsection*{\DLC is fast enough for future real-time tracking}
Because it relies on an artificial neural network, the \DLC model finds  correspondence for a set of neurons faster than traditional methods (Table \ref{table:timing}). 
From the time that a configuration  of  segmented neurons is loaded onto a GPU, it takes only an average of 10 ms for the \DLC model to label all  neurons  on  a  2.4 GhZ Intel machine with an NVIDIA Tesla P100 GPU. If not using a GPU, the model labels neurons  in 50 ms.  In contrast, on the same hardware it takes CPD  930 ms  and it takes NeRVE on average  over  10 seconds.  
The \DLC model may be  a good candidate for potential closed-loop tracking applications because its speed of 100 volumes per second is an order of magnitude faster than the 6 to 10 volumes per second recording rate typically used in whole-brain imaging of freely moving \textit{C. elegans}~\cite{nguyen_whole-brain_2016, venkatachalam_pan-neuronal_2016}.  We note that for a complete closed-loop tracking system, fast segmentation algorithms will also be needed in addition to the fast registration and labeling algorithms presented here. The \DLC model is agnostic to the details of the segmentation algorithm so it is well suited to take advantage of fast segmentation algorithms when they are developed.

\begin{table}[!ht]
\centering
\caption{
{\bf Time required to find neural correspondence}}
\begin{tabular}{c|c}
\bf Method & \bf Time(s/Volume)\\ \hline
CPD~\cite{myronenko_point_2010} & 0.93 \\
NeRVE(1) ~\cite{nguyenAutomaticallyTrackingNeurons2017}& $10$\\
NeRVE(100) ~\cite{nguyenAutomaticallyTrackingNeurons2017}& $>10$\\
\DLC [this work] & \bf 0.01\\
\end{tabular}
\begin{flushleft}Table shows the measured time  per volume required for  different models to predict neural correspondence of  a single volume. Time required  is measured after neuron segmentation is complete and a configuration of neural positions has been loaded into memory. The same hardware is used for all models.
\end{flushleft}
\label{table:timing}
\end{table}

The \DLC model uses  built-in libraries to parallelize the computations for labeling a single volume, and this contributes to its speed.
In particular, each layer of the neural network contains thousands of artificial neurons performing the same computation. 
Computations for each neuron in a layer can all be performed in parallel and modern  GPUs have as many as  3,500 CUDA cores. 

In practice, the method is even faster for post-processing applications (not-realtime) because it is also parallelizable at the level of each volume. Labeling one volume has no dependencies on any previous volumes and therefore each volume can be processed simultaneously. The number of volumes to be processed in parallel is limited only  by number the of volumes that can be loaded onto the memory of a GPU. When tracking during post-processing in this work, we used 32 volumes simultaneously.

\subsection*{\DLC accurately finds neural correspondence  across individuals}

\begin{figure}[!htpb]
 	\includegraphics[width=0.9\linewidth]{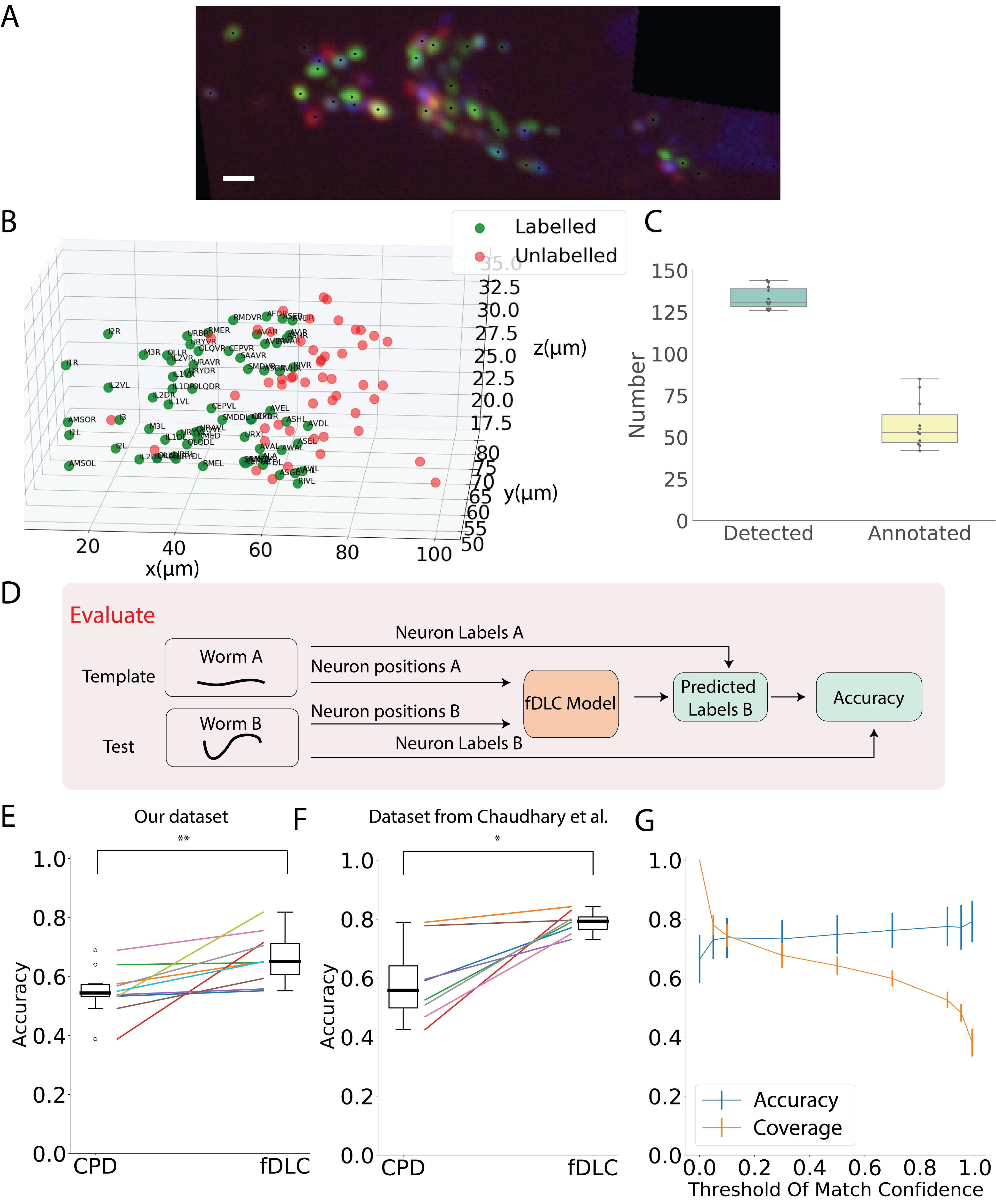}
\caption{{\bf \DLC model  finds neural correspondence across individuals.}
(A)   Fluorescence image shows neuronal nuclei of a NeuroPAL worm. A single optical slice  is shown from an optical stack. (Scale bar, $10\mu m$). Genetically encoded color labels in Neuropal animals aid ground truth manual neural identification \cite{yeminiNeuroPALMulticolorAtlas2020} and are used here to evaluate performance.
Black dots indicate  neurons found via automatic segmentation.
(B) Locations of all segmented  neurons from A. Neurons that additionally have  a human annotated label are shown in green. Those that a human was unable to label are red.   (C) Number of segmented neurons (mean 133.6) and  subset of those that were given human annotations (mean 57.5) is shown for  11 NeuroPAL individuals.  Box plot shows median and interquartile range. 
(D) Pipeline to evaluate \DLC model  performance across NeuroPAL individual is shown. Predicted labels are compared with human annotated labels to compute accuracy.
(E) Performance of the \DLC model and CPD is shown evaluated on NeuroPAL recordings using position information alone. Accuracy is the fraction of labeled neurons present in both test and template that are correctly matched.   Performance is evaluated on 10 pairs of 11 recordings, where the template is always the same (Worm A).    ($p=0.005$,  Wilcoxon signed-rank test). (F) Performance evaluated on a separate publicly accessible  dataset of 10 NeuroPAL individuals from \cite{chaudharyAutomatedAnnotationCell2020} ($p=0.012$,  Wilcoxon signed-rank test). (G) Accuracy and fraction of neurons matched (coverage) is shown as a function of the minimum confidence needed to accept a match. Same reccording as in E. By default, no threshold is applied.    E and F use no threshold. }
\label{fig4}
\end{figure}

 Having shown that \DLC performs well at identifying neurons within the same individual, we wanted to address its capability to identify neurons across different animals. Identifying corresponding neurons across individuals is crucial for studying the nervous system.  However, finding neural correspondence across individuals is  more challenging than  within an individual because  there is variability in neuronal position from both the animal's movement as well as from development.  To evaluate the \DLC model's performance at finding neural correspondence across individuals,  we applied the same synthetically-trained \DLC model to a set of 11 manually annotated NeuroPAL worms. 
 
 NeuroPAL worms contain  genetically encoded multicolor fluorescently labeled neural landmarks to aid neural identification  ~\cite{yeminiNeuroPALMulticolorAtlas2020}.  For each of the 11 Neuropal recordings, neurons were automatically segmented and manually annotated  based on the neuron's position and color features as described in ~\cite{yeminiNeuroPALMulticolorAtlas2020} (see Fig.~\ref{fig4}A,B). Across the 11 animals, a human  assigned a ground-truth label to a mean  of 43\% of segmented head neurons, providing approximately 58 labeled neurons per animal (Fig.~\ref{fig4}C). The remaining neurons were not confidently labeled and thus not included in this analysis.  We selected as template the recording that contained the largest of confidently labeled human annotated neurons. We evaluated our model by comparing its predicted correspondence between neurons in the other 10 test datasets and this template, using only position information. All 11 ground-truth recordings were held-out in that they were not involved in the generation of the synthetic data  that had been used to train the model.

We applied the synthetically-trained \DLC model  to each pair of held-out NeuroPAL test and template recordings and calculated the accuracy as the fraction of labeled neurons present in both the test and the template that was correctly matched.  Across the 10 pairs of NeuroPAL recordings using position information alone, the \DLC model had an accuracy of 65.8\%, significantly higher than  the CPD method's accuracy of 54.7\% ($p=0.005$,  Wilcoxon signed-rank test).   

For each neuron $i$ in the test recording, the \DLC model returns the  probability with which that neuron corresponds to each possible neuron $j$ in the template, $p_{ij}$. We wondered whether we could better use the likelihood information about potential matches generated  by the algorithm. A Hungarian algorithm finds the most probable match by considering all $p_{ij}$s for all neurons in the test. By default  we use this best match in evaluating performance. But the  $p_{ij}$s also provide the user with a  list of  alternative matches ranked by their likelihood. We  therefore also assessed the accuracy for the top 3 most likely candidate neurons. In this context, we define accuracy as  the fraction of neurons for which the correct match is included in the list of the top 3 candidates. As before, the denominator is taken to be the number of ground-truth labeled neurons that appear in both test and template. When considering the top 3 neurons,  the \DLC model achieves an accuracy of 84.4\% using only position information.

\subsubsection*{Validating on an alternative dataset}
Data quality, selection criteria,  human annotation,  hardware  and  preprocessing can all vary from lab-to-lab making it challenging to directly compare methods. To validate our model against different measurement conditions and to allow for a direct comparison with another recent method, we applied our \DLC model to a previously published  dataset of 9 NeuroPAL individuals ~\cite{chaudharyAutomatedAnnotationCell2020} (Fig.~\ref{fig4}F). This public dataset used different imaging hardware and conditions and was annotated by human experts from a different group.  On this public dataset,  our method achieved 78.9\% accuracy while CPD achieved  59.4\%. When assessing the top 3 candidate accuracy, the \DLC model performance was 91.3\%. The \DLC model performance was overall  higher on the published dataset compared to our newly collected dataset  presented here. This suggests that  our method performs well when applied to real-world datasets in the literature.

\begin{table}[!htbp]
\centering
\caption{
{\bf Comparison of model performance on additional dataset}}
\begin{tabular}{l|ccc}
\bf Method & \bf Accuracy & $N$  & Reported in\\ \hline
CPD& 59\% &8 & This work\\
CRF (open atlas) & $\approx 40$\% &9 & \cite{chaudharyAutomatedAnnotationCell2020}\\
CRF (data driven atlas) & 74\% &9 & \cite{chaudharyAutomatedAnnotationCell2020}\\ 
\DLC & \bf 79\% &8&This work \\
\end{tabular}
\begin{flushleft} Table lists  reported mean accuracy  of different models evaluated on the same publicly accessible dataset from \cite{chaudharyAutomatedAnnotationCell2020}. $N$ indicates the number of template-test pairs used to calculate accuracy. (CRF method uses an atlas as the template, whereas we randomly take one of the 9 individuals and designate that as the template).  CPD and \DLC  performance on this dataset are also shown in Fig.~\ref{fig4}F.  
\end{flushleft}
\label{table2}
\end{table}

We further compared the \DLC model to a recent model called Conditional Random Fields (CRF) from \cite{chaudharyAutomatedAnnotationCell2020}. We compared the  reported performance of the CRF model on the published dataset to the performance of the \DLC model evaluated on the same dataset (Table \ref{table2}). The CRF model has two variants an ``open atlas'' and a ``data-driven atlas.'' \DLC accuracy is notably higher than the reported CRF performance for either variant, although we are unable to test for statistical significance.  We note the \DLC method also offers other advantages compared to the CRF approach in that the \DLC method is  optimized for speed and avoids the need to transform the worm into a canonical coordinate system.  Taken together, we conclude that the \DLC model's accuracy compares favorably to that of the  CRF model while also providing other advantages.


\subsubsection*{Tradeoff between performance and coverage}

Our model predicts correspondence for all pairs of neurons, but the confidence of the predicted label varies by pair. Therefore, there is an inherent tradeoff between the accuracy of matched neurons (performance) and the fraction of neurons matched (coverage).The probability $p_{ij}$ in Equation \ref{eq:Pij} (in Materials and methods) serves as an estimate of the confidence level. By default the \DLC model matches all neurons regardless of confidence and, with the exception of Fig.~\ref{fig3}, that is how performance is  evaluated in this work. (In Fig.~\ref{fig3} a minimum confidence was imposed  to better compare to the NeRVE approach, which also only matches  a subset of neurons. Details are  described in methods section ``Minimum confidence threshold.'')

In practice, it is often desirable to focus on only those neurons that are matched with very high confidence. We therefore tried restricting the model to only include matches above certain confidence thresholds. As the confidence threshold increased, model accuracy  increased at the cost of the number of labeled neurons (Fig.~\ref{fig4}G). For a threshold of $p=0.05$ we observed an increase in the accuracy on our Neuropal dataset from 65.8\% to 73.0\% while only reducing coverage from 100\% to 78.0\% of the total neuron matches (intersection of test and template). By increasing the threshold to 0.99, the accuracy increases to  79.2\% but now only 38.1\% of the neurons are matched. This feature of the model allows the experimenter to transparently tune the trade-off between  prediction accuracy and coverage.

\subsection*{Incorporating color information}
\begin{figure}[!htbp]
 	\includegraphics[width=0.9\linewidth]{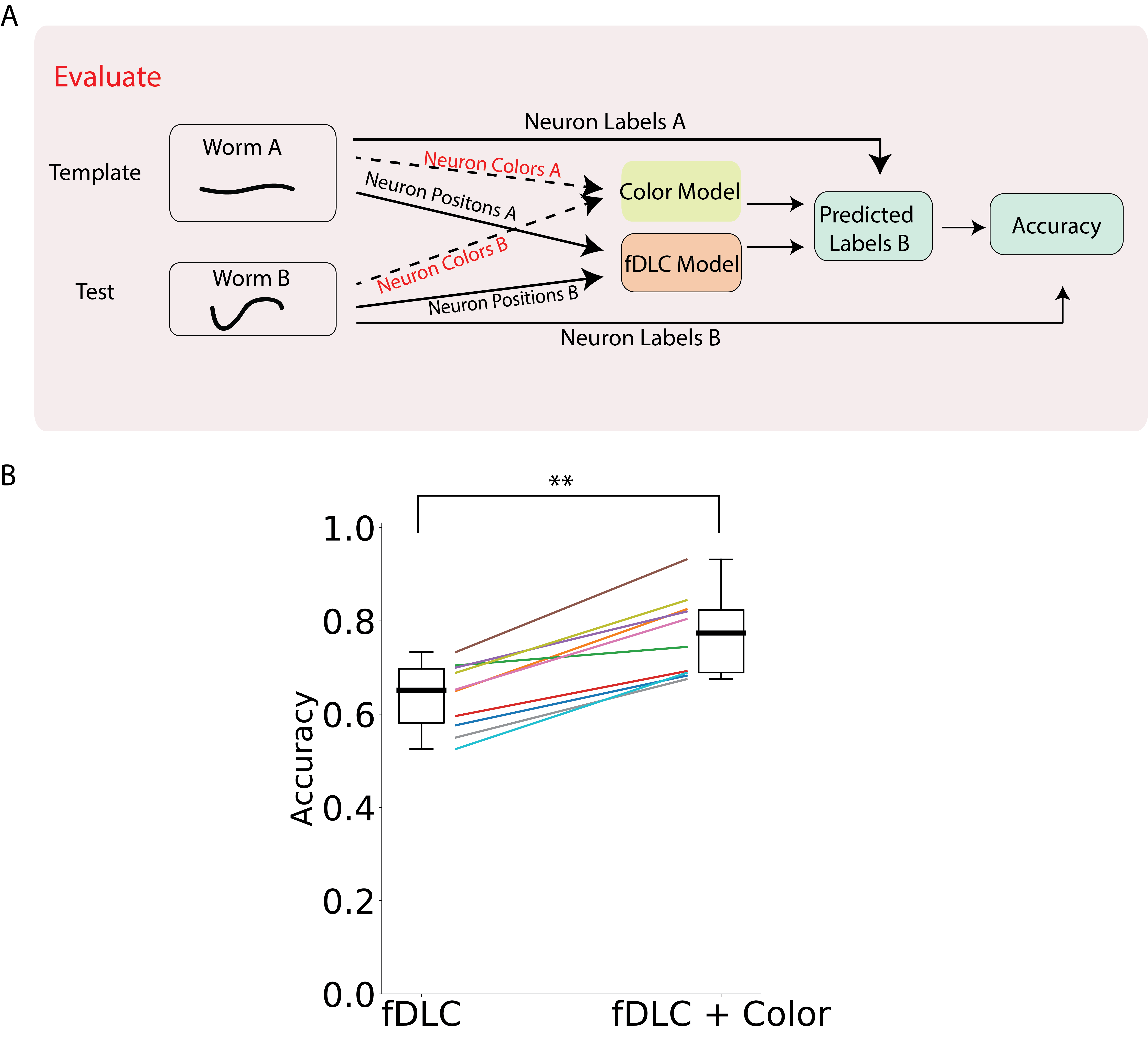}
\caption{{\bf  \DLC performance when incorporating  color features}
(A)  Pipeline to evaluate \DLC  performance across animals with additional color features. A simple color model is added in parallel to the \DLC model to use both  color and position information from  11 NeuroPAL recordings.  Accuracy is calculated from ground truth human annotation and is the fraction of labeled neurons present in both test and template that are correctly matched . Matching probabilities from the color and \DLC  models are combined to form the final matching probabilities.
(B) Accuracy of the  position-only \DLC model  and the combined \DLC and color model are evaluated on  11 NeuroPAL recordings (same recordings as in Fig.~\ref{fig4}).  $p=\num{5.0e-3}$, Wilcoxon signed rank test.  }
\label{fig5}
\end{figure}

Our method only takes positional information as input to predict neural correspondence. However, when additional features are available, the position-based predictions from the  \DLC model can be combined with predictions based on other features to improve overall performance. As demonstrated in \cite{yeminiNeuroPALMulticolorAtlas2020}, adding color features from a NeuroPAL strain can reduce the ambiguity of predicting neural correspondence. 
We applied a very simple color model  to calculate the similarity of color features between neuron $i$ in the test recording to every possible neuron $j$ in the template. The color model returns matching probabilities, $p_{ij}^\text{c}$ based on the Kullback-Liebler divergence of the normalized color spectra in a pair of candidate neurons (details described in Materials and methods). The color model  is run in parallel to the \DLC model (Fig.~\ref{fig5}A). Overall matching probabilities $p_{ij}^\text{all}$ that incorporate both color and position information are calculated by combining the color matching probabilities $p_{ij}^\text{c}$ with the position probabilities $p_{ij}$.  The Hungarian algorithm is run on the combined matching algorithm to predict the best matches. 

Adding color information increased the \DLC model's accuracy from 65.8\% to 76.5\% (Fig.~\ref{fig5}B) when evaluated on our dataset, and improved the accuracy in every recording pair.   The top 3 candidate labels attained 92.4\% accuracy. Accuracy was calculated from a comparison to human ground truth labeling, as before. 

We chose a trivially simple color model in part to demonstrate  the flexibility with which the \DLC model framework can integrate information about other features.
Since our simple color model  utilized no prior knowledge about the distributions of colors in the worm, we would expect a more sophisticated color model, for example, the statistical model used in \cite{yeminiNeuroPALMulticolorAtlas2020}, to do better. And indeed that model evaluated on a different dataset is reported to have a  higher performance with color than our model on our dataset (86\% reported accuracy in \cite{yeminiNeuroPALMulticolorAtlas2020} compared to 77\% for the \DLC evaluated here). But that model also performs much worse  than  \DLC  when both are restricted to use only neural position information (50\% reported accuracy for \cite{yeminiNeuroPALMulticolorAtlas2020} compared to 66\% for the \DLC).  
Together, this suggests the \DLC model framework can take advantage of additional feature information like color and still perform  well when such information is missing.

\section*{Discussion}
The \DLC model finds neural correspondence within and across individuals with an accuracy that compares favorably to other methods.  
The model focuses primarily on identifying neural correspondence using   position information alone. The \DLC model  framework also makes it easy to  integrate other features. We demonstrated that color information could be added by  integrating the \DLC model with a simple color model to increase overall accuracy. We expect that performance would improve further with a more sophisticated color model that takes into account the statistics of the  colors in a  NeuroPAL worm \cite{yeminiNeuroPALMulticolorAtlas2020}. 

The \DLC model framework offers a number of additional advantages beyond accuracy. First, it is versatile and general. The same pre-trained model performed well at both tracking neurons within a freely moving individual across time and at  finding neural correspondence across different individuals. It achieved even higher accuracy on a publicly accessible dataset acquired on different hardware with different imaging conditions from a different group.   This suggests that the framework should be applicable to many real-world datasets.

Second, the model reports the relative confidence in its  estimate of each neuron's correspondence. An experimenter can therefore tune the overall accuracy of the model's predictions, at the cost of leaving some neurons unmatched, by simply setting a desired minimum confidence.   Similarly, the model  also provides probability estimates of all possible  matches for each neuron, not just the most likely.  This allows an experimenter to  consider a collection of   possible matches  such as the top 3.

In contrast to previous methods,  an advantage of the \DLC method is that it  does not require the worm to be straightened, axis aligned, or otherwise transformed into a  canonical  coordinate system. This eliminates an  error-prone and often manual step.  Instead, the \DLC model  finds neural correspondence directly from neural position information even in worms that are in different poses or orientations.

Importantly, the model is trained entirely on synthetic data, which avoids the need for large experimentally acquired ground truth datasets to train the artificial neural network.  Acquiring ground truth  neural correspondence in \textit{C. elegans} is  time consuming, error prone, and often  requires manual hand annotation. The ability to train the \DLC model with synthetic data derived from measurements  alleviates this bottleneck and makes the model attractive for use with other organisms with stereotyped nervous systems where ground truth datasets are similarly challenging to acquire.

The model is also  fast and  finds neural correspondence of a new neural configuration in 10 ms. This speed is sufficient to keep up with real-time calcium imaging applications and will be valuable for  future efforts combining large scale population calcium imaging with   closed loop targeted optogenetic stimulation in freely moving animals  \cite{leiferOptogeneticManipulationNeural2011, stirmanRealtimeMultimodalOptical2011, kocabasControllingInterneuronActivity2012, shipleySimultaneousOptogeneticManipulation2014}.  We note that to be used in a real-time closed loop application, our \DLC model would need to be combined with faster segmentation algorithms because current segmentation algorithms are too slow for real-time use. Because segmentation can  be easily paralellized, we  expect that faster segmentation algorithms will be developed soon. 

Many of the advantages listed here stem from the \DLC model's use of  the transformer  architecture \cite{bibTransformer} in combination with supervised learning. The transformer architecture, with its origins in natural language processing, is well suited to find spatial relationships within a configuration of neurons. By using supervised learning on empirically-derived synthetic training data of animals in a variety of different poses and orientations, the model is forced to   learn relative spatial features within the neurons that are informative for finding neural correspondence across  many postures and conditions.  Finally, the transformer architecture leverages recent advances in GPU parallel processing for speed and efficiency, which will help pave the way for real-time, closed-loop optogenetic experiments in freely moving worms.

\section*{Acknowledgments}
We thank Ev Yemini and Oliver Hobert of Columbia University for strain OH15262.  We acknowledge productive discussions with John Murray of University of Pennsylvania.  This work used computing resources from the  Princeton Institute for Computational Science and Engineering. Research reported in this work was supported  by the Simons Foundation under awards  SCGB \#543003 to AML and SCGB \#697092 to SWL;  by the National Science Foundation, through an NSF CAREER Award to AML (IOS-1845137) and through the Center for the Physics of Biological Function (PHY-1734030); and by the National Institute of Neurological Disorders and Stroke of the National Institutes of Health under award numbers R21NS101629 to AML and 1R01NS113119 to SWL.

\section*{Additional information}
\subsection*{Author contributions}
Xinwei Yu, Conceptualization, Software, Formal analysis, Investigation, Methodology, Visualization, Validation, Writing-original draft, Writing—review and editing; Matthew S. Creamer, Investigation, Collected Data, Writing—review
and editing; Francesco Randi, Resources, Designed optics and related software libraries, Writing—review and editing; Anuj Sharma, Resources, Writing—review
and editing, Performed all transgenics; Scott W. Linderman, Conceptualization, Funding
acquisition, Writing—review and editing; Andrew M Leifer, Conceptualization, Supervision, Funding
acquisition, Writing—original draft, Project administration, Writing—review and editing
\nolinenumbers

%
%
%

\bibliography{references}

\end{document}